# Synthesis and Tailored Properties Towards Designer Covalent Organic Framework Thin Films and Heterostructures

Lucas K. Beagle[†1,2], Qiyi Fang[†3], Ly D. Tran[1,2], Luke A. Baldwin[1], Christopher Muratore[4], Jun Lou*[3], Nicholas R. Glavin*[1]

[1]Materials and Manufacturing Directorate, Air Force Research Laboratory, Wright-Patterson AFB, OH 45433, USA
[2]UES, Inc., Beavercreek, OH 45431, USA
[3]Department of Materials Science and Nanoengineering, Rice University, Houston, TX 77005, USA
[4]Department of Chemical and Materials Engineering, University of Dayton, Dayton, OH 45469, USA

Co-corresponding authors: jlou@rice.edu, nicholas.glavin.1@us.af.mil

**Abstract**

Porous polymeric covalent organic frameworks (COFs) have been under intense synthetic investigation with over 100 unique structural motifs known. In order to realize the true potential of these materials, converting the powders into thin films with strict control of thickness and morphology is necessary and accomplished through techniques including interfacial synthesis, chemical exfoliation and mechanical delamination. Recent progress in the construction and tailored properties of thin film COFs are highlighted in this review, addressing mechanical properties as well as application-focused properties in filtration, electronics, sensors, electrochemical, magnetics, optoelectronics and beyond. Additionally, heterogeneous integration of these thin films with other inorganic and organic materials is discussed, revealing exciting opportunities to integrate COF thin films with other state of the art material and device systems.

1. Introduction

The past few decades have experienced revolutionary advances in nanomaterials and nanoscience, with tailored properties realized from manipulation of materials at size-scales of 100 nm and below.[1,2] An emerging class of organic nanomaterials referred to as covalent organic frameworks (COFs) exhibit controllable nanoscale porosity in a highly ordered, modular framework structure.[3–5] The reticular network, much like related metal organic frameworks (MOFs), allows for nearly infinite combinations of precursors to be used to in its construction, with integration of different linkers into the same framework resulting in multifunctional properties within one material.[6–8] The electron rich π-systems in 2D COFs demonstrate tunable band gaps, strong interactions with other molecules and materials, and unique optical and electronic properties.[9,10]

The materials science community has increasing need to explore new smart, reconfigurable and responsive materials often which include multiple, differing properties to accomplish a wide array of tasks.[11–16] Utilizing the intrinsic nature of COFs, one material can have disparate properties expressed simultaneously within the framework.[11] Incorporation of these COFs into thin films increases their usefulness by accentuating their properties and allows integration into modern catalytic, electrical, optical, magnetic, and filtration constructs and devices (Figure 1). These device constructs are only accessible due to the synergistic multifunctionality afforded by COFs, especially thin films and their heterogeneous integration with other substrates and 2D materials.[17–20] Thin films with controllable thicknesses from 500 nm down to a few or monolayers demonstrate properties that may not be available in the bulk material due to layer alignment effects and quantum confinement. COFs of nanoscale-thickness can be integrated into devices including field effect transistors (FETs),[21,22] photodetectors,[23–25] filtration apparatii,[26–28] delivery systems[29,30] and solid-phase catalyst and supports.[31,32]

Innovations in smart materials are an increasing necessity in our efforts to develop better functional surfaces especially in the application area of electronics, mechanical, spintronics and optics. Response to stimuli that have included the presence of moisture, pH, chemisorption, electric field, surface potential, and photons[33–38] have been utilized in the conveyance of information to a logic device initiating the subsequent signal cascade for detection.[39] Through incorporation of various functional groups within the framework, initiation of chemical reactions can sense the presence of chemical analytes and even further reactions will act sacrificially in the framework during the process.[40] When the framework undergoes chemical changes the optical and electronical properties are also modified, often resulting in changes of measurable significance.[41] The tunable band gap demonstrated by the semi-conductor nature of certain COF allow for the sensing of surface potential and electric fields.[24] Due to the inherent modularity of linkers and vertices in thin film COFs there is an intrinsic built-in tunable optical bandgap with choice of the primary structure. At the device level the need for thin film COFs to be used in detection of analytes, the presence or absence of light, sequestration or filtration of analytes and release of targets upon stimulation are of importance to the community.[42–44]

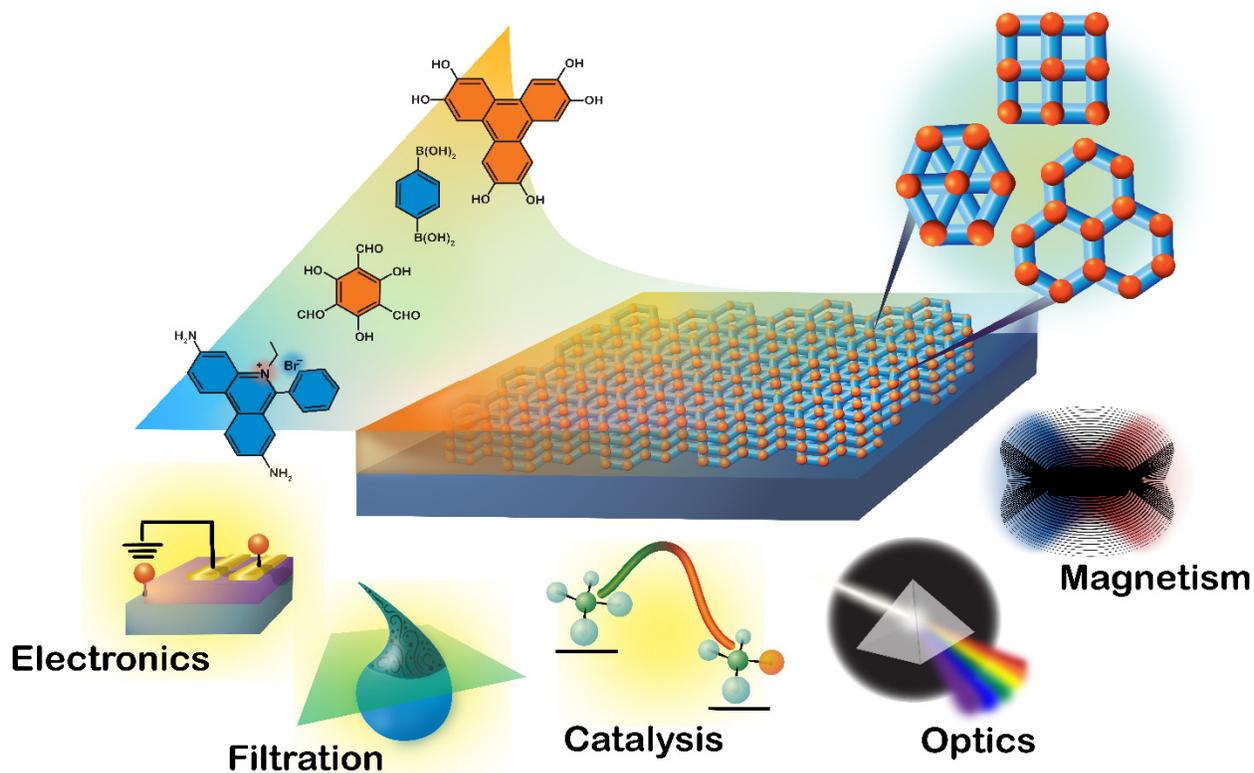

**Figure 1**: Thin film COF schematic depictions including various applications.

The chemical and physical properties of COFs rely on selection of linker and vertex molecules that comprise a molecular framework with controllable pore sizes and shapes (Figure 2).[16] The overarching linker-vertex chemistries result from the reversible reaction which covalently connects the individual linkers and vertices. These reversible chemical reactions include boronate ester COFs from the reaction of boronic acids and alcohols or imine formation though a condensation of aldehydes and amines.[45] These dynamic bonds, such as imine bonds, can even be further reduced to a less hydrolysable amide bond.[46,47] Highly irreversible reactions can also be used to form the linkages, in which thermodynamically stable products from C-C bonds are formed through transition metal cross-coupling reactions and other cascade reactions.[48,49] Each linkage modality strongly influences the COF function and stability, resulting in certain linkages being favorable to resisting acid or solvent susceptibility, pore collapse, and temperature breakdown.[4]

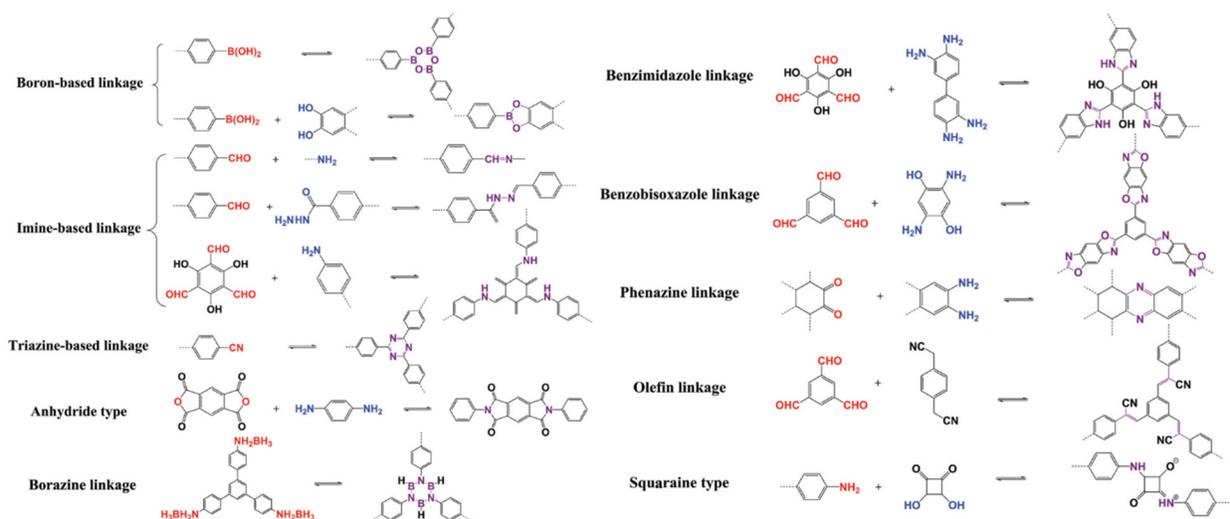

**Figure 2**: Example covalent organic frameworks linker chemistries. Reproduced with permission from Ref [16] Copyright 2019, The Royal Society of Chemistry.

Pore geometry is one of the most easily tailored properties in COFs through selection of monomer linker and vertex structure (Figure 3a).[50] The simplest linkers involve two reactive moieties whereas more complex linkers can have greater numbers of sites to increase the linking attachments. In pore shape construction, the choice of vertex is the most vital, with attachment moieties than their linker counterparts for control of the COF pore architecture from hexagonal to rhombus, square, or triangular. With this linker/vertex combination, a single two attachment linker combined with multiple different vertices may form several different pore shapes within one structure, such as a mixed triangular and hexagonal structure (Figure 3b).[25] COF synthesis routes can also take advantage of an inverse design paradigm, in which the chosen pore shape determines which class of linkers and vertices are selected.[10]

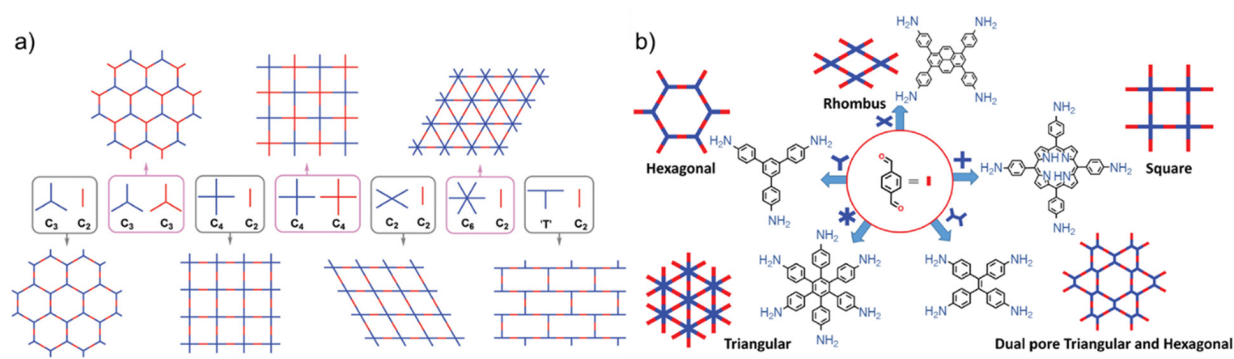

**Figure 3**: Engineering of COF pore geometry via selection of linker and vertex a) Reproduced with permission from Ref [50] Copyright 2020, The Royal Society of Chemistry. b) imine based linkages showing pore shapes achieved by varying the vertex with the same linker monomer. Reproduced with permission from Ref [25] Copyright 2017, WILEY-VCH Verlag GmbH & Co. KGaA, https://doi.org/10.1002/aenm.201700387.

Typically, the nanoporous COFs are synthesized as insoluble powders via solvothermal,[3] ionothermal,[49] sonochemical,[51] microwave-assisted,[52] and mechanochemical[53] approaches. More recently, significant efforts have been made to synthesize mono- and multi-layer COF thin films as well as COF nanosheets via solvothermal, interfacial, vapor-assisted synthesis and chemical/mechanical exfoliation methods,[17,18] largely driven by many promising applications enabled by the more precise thin film control. Inspired by many important opportunities offered in this highly promising yet under-developed research space, this review first focuses on COF thin films prepared by interfacial synthesis methods which offer great flexibility for constructing heterostructures while maintaining high crystalline quality important for device applications. This is followed by carefully assessing the oftentimes overlooked but critical mechanical properties of COF thin films. Next, application-focused properties in filtration, electronics, sensors, electrochemical and magnetic devices, optoelectronics and beyond are discussed before highlighting the emerging opportunities of heterogeneous integration of COF thin films with other material systems.

## 2. COF Thin Film Preparation and Synthesis

The fundamental interfacial synthesis methods to synthesize COF thin films can be broken down into three specific methodologies referred to as solid-liquid, liquid-liquid and gas-liquid based approaches, as illustrated below in Figure 4. As common characteristic of COF powders is low solubility, requiring tailoring of these processes to overcome this challenge to form high quality COF thin films. In the following section, we will review the latest progress made in developing these synthesis methods. It is worth noting that COF thin films synthesized by interfacial approaches generally have tunable thicknesses from mono- to multi-layers, well-oriented and highly crystalline, and can be on different substrates or free-standing and easily transferrable, making them well-suited for building advanced devices and systems as well as constructing multi-functional heterostructures.

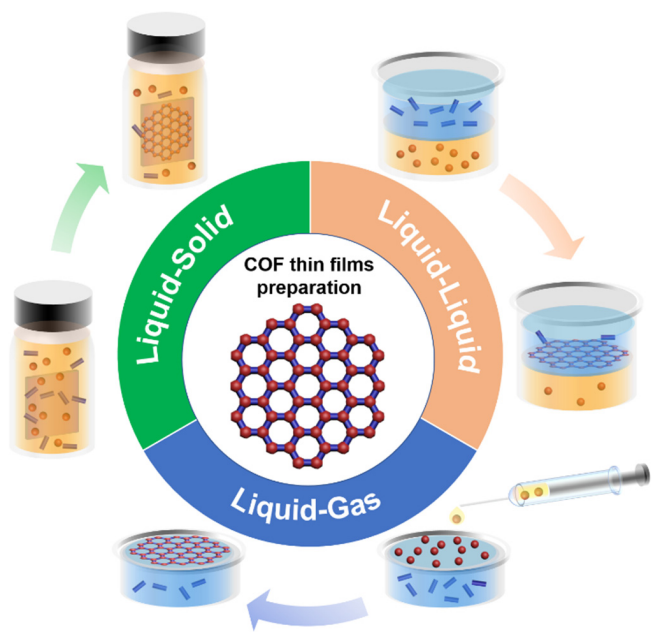

**Figure 4**: Schematic illustration of three major synthetic routes to prepare COF thin films by interfacial methods.

Preparation of COF thin films at the solid/liquid interface

Pioneering work in the solid/liquid interface preparation method by Dichtel and co-authors demonstrated that oriented COF-5 thin films form on single layer graphene (SLG) substrates by immersing a SLG/Cu substrate into the solution of 1,4-dioxane and mesitylene containing monomers (HHTP and PBBA) at 90°C (Figure 5a).[54] Grazing incidence diffraction (GID) showed that the hexagonal lattice plane was parallel to the substrate surface (Figure 5b). As this method involved the diffusion and organization of monomers from liquid to solid substrates, it was assumed that the origin of the preferred orientation is due to the strong π-π interaction between monomers and substrates like graphene. However, Bein et al. successfully prepared COF films by co-condensation of the benzo[1,2-b:4,5-b']dithiophene-2,6-diyldiboronic acid (BDTBA) and polyol 2,3,6,7,10,11-hexahydroxytriphenylene (HHTP) on inorganic polycrystalline substrates such as indium-doped tin oxide (ITO)- or NIO/ ITO-coated glass substrates.[55] The GID result of the concentrated reflection near q(z)=0 verified that the as-prepared COF films were highly oriented in this case, implying that the orientation of the film might not be strongly correlated to the π-π interactions.

Oriented thin films synthesized from imine-based COF can also be prepared in a similar fashion, where Liu et al showed the co-condensation of tetrathiafulvalene tetraaldehyde and 1,4-diamidebenzene on transparent ITO substrates.[9] Grazing incident small angle x-ray scatting (GIWAXS) characterization of the COF film revealed the polycrystalline nature with preferential orientation of the columns normal to the substrate. In follow-up studies, various functional oriented thin films of imine-based COF were prepared and were demonstrated to have high conductivity,[12,56,57] improved photo responsibility,[23,58,59] catalytic[60–62] and sensing properties.[34,63,64]

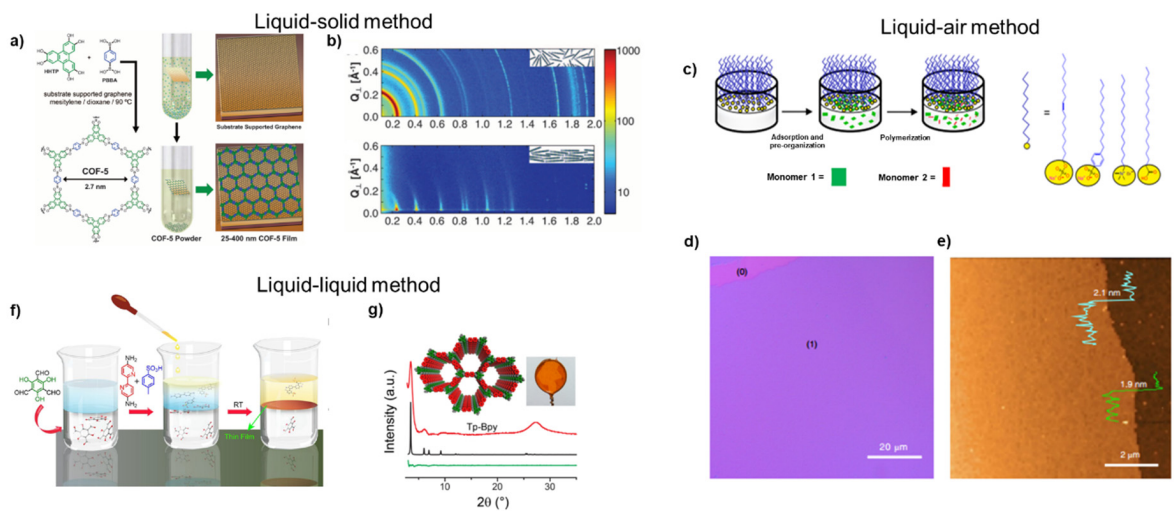

**Figure 5**: Thin film COF synthesis highlighting the liquid-solid, liquid-liquid and liquid-air interfacial methodologies. a) Schematic illustration of the solvothermal synthesis of COF-5 thin films b) X-ray scattering data from COF-5 powder (inset: randomly oriented COF-5 grains in the powder); GID data from a COF-5 film on SLG/Cu (inset: oriented COF-5 grains in the film). Reproduced with permission from Ref [54]. Copyright 2011, American Association for the Advancement of Science c) Schematic illustration of the preparation of COF films on water surface assisted by surfactant d) optical image of the COF thin film e) An AFM image of the COF thin film. Reproduced with permission from Ref [65] Copyright 2019, Springer Nature. f) schematic illustration of interfacial polymerization process used to prepare COF thin films g) Experimental powder X-ray diffraction pattern of COF film (inset: freestanding thin film transferred on a wire loop). Reproduced with permission from Ref [66].Copyright 2017, American Chemical Society.

Preparation of COF thin films at the air/liquid interface

The air/liquid interface method shows great potential for the preparation of single layer materials, in which hydrophobic monomers are confined between air and liquid interfaces and spontaneously form a single layer upon reaction.[67] In 2014, Schlüter et al. first reported monolayer two-dimensional polymer (2DP) prepared by Langmuir-Blodgett trough (LB trough).[68] The thickness of this covalent monolayer was measured to be around 1.4 nm and can be transferred onto holey substrates to form a free-standing film, suggesting good film rigidity. Although several single layer 2DPs and MOFs prepared from LB trough have been reported,[69–72] the synthesis of COF monolayer from this method remains challenging. In one pioneering study in 2016 by Dai et al, imine-based covalent monolayer was synthesized by LB trough methodology, specifically, the two monomers: terephtharaldehyde (TA) and 1,3,5-trihexyl-2,4,6-tris(4-aminophenyl)benzene (Hex-TAPB) were allowed to react at air/water interface to yield polyimine monolayer overnight. While the thickness was measured to be 0.7 nm, corresponding to monolayer thickness, the materials have have low mechanical strength and suffer from rupture under low applied stresses. While Raman spectra showed the successful condensation of the aldehyde and amide groups, the internal long-range ordering had not been proven.[73] However, in 2016, Feng et al prepared wafer-size porphyrin-containing monolayer 2DP from LB trough in which the monolayer thin film showed long-range ordering.[74]

Although mono- or multilayer COF films prepared from LB trough have been reported,[26,75] it is not an ideal method for scalability and reproducibility as it requires the use of a very delicate instrument. In addition to the LB trough method, ultrathin films can be prepared by the spread of a small amount of organic precursor solution on the water surface.[21] After evaporation of organic solvent, an ultrathin COF film was seen on the water surface, but the thickness of the film was not easily controllable. To solve this problem, Choi et al. reported large-area synthesis of a layer-controlled COF film, even down to monolayer, using a photon-assisted imine condensation on the water interface.[76] Using a solution of optimized polarity, the precursor solution is spread across the water surface, this allows for a uniform distribution of the monomer layers across the water. The thickness of the film in turn is controlled by the concentration of the precursor solution. Using this method, Choi and coworkers successfully prepared mono- to octa-layer COF films. Using a surfactant, Feng et al. developed a novel approach in the preparation of ultrathin COF films.[65] By means of a self-assembled surfactant monolayer on the water surface (Figure 5c), the monomer would be absorbed on the liquid/air surface facilitated through interaction with the surfactant. The pre-organized monomer layers react with subsequent monomers to form crystalline COF thin films. Furthermore, they were able to prepare large area films (Figure 5d) with a thickness of around 2 nm (Figure 5e), with long range order being confirmed by TEM. It was noted that the addition of surfactant was the key to preparation of the crystalline films as it promoted organization of the monomers. Without surfactant, only amorphous films were obtained. Further, this method was applied to prepare imide, imine and boronate ester COF films.[77,78]

Preparation of COF thin films at the liquid/liquid interface

Liquid/liquid interface-based method, also known as the interfacial polycondensation, has been widely applied for preparation of polymers, such as polyamides and polyesters in bulk scale.[79–81] Using this method, two monomers are dissolved in two immiscible solvents, usually water and organic solvent, respectively, and the reactions between two monomers are restricted to the liquid/liquid interface.[66] Compared to solid/liquid and air/liquid interface based methods, this method is more flexible and scalable. Focusing on imine-based COF, the rapid formation of the Schiff base typically leads to amorphous polymers, for the preparation of long range ordered polymer films exercising control over reaction rate to increase crystallinity is crucial. Banerjee et al. applied a salt-mediated technique by adding p-Toluenesulfonic acid (PTSA), upon addition the amide monomers form amide-PTSA salt and dissolve in water (Figure 5f). A series of 1,3,5-trisformylphoroglucinol (Tp) COF thin films including Tp-BPy (Bpy: 2,2'-bipyridine-5,5'-diamine), Tp-Azo (Azo: 4,4'-azodiamine), Tp-Ttba (Ttba: 4,4',4"-(1,3,5-triazine-2,4,6-triyl) tris(1,1'-biphenyl)trianiline), Tp-Tta (Tta: 4,4',4"-(1,3,5-triazine-2,4,6-triyl)trianiline) were synthesized at the liquid-liquid interface of DCM/water system. Compared to free amines in the water phase, the amide-PTSA salt has a decreased diffusion through the interphase and thereby increased crystallinity of the polymer film. Applying this method, these COF films with thickness ranging from 50 to 200 nm, and their long-range ordering confirmed by XRD characterizations (Figure 5g). These self-standing thin films showed moderate mechanical robustness as the films were stable up to several weeks after being transferred or mounted on a U-shaped glass loop. These successes soon sparked a series of studies where several COF thin films were prepared and applied in many areas.[15,82–85] However, to improve the crystallinity, this method usually requires long reaction time. To solve this issue, Dichtel et al. introduced high performance Lewis acid, Scandium triflate (Sc(OTf)3), as catalyst.[86] The Sc(OTf)3-catalyzed polymerization occurs rapidly at room temperature. In this method, they dissolved both monomers in organic phase, and added catalyst in the water phase to prepare highly crystalline COF thin films at the interface within 30 minutes.[87] The thicknesses of the COF films could be controlled by the concentrations of monomers, ranging from 2 nm to few microns. These films could be transferred to arbitrary supporting substrates for further applications.

Interestingly, another area of control was achieved by controlling the dimensionality of the water phase by spreading a thin film of it across a super hydrophobic gel.[88] Wang et al. explored a tri-phase system, in which the ultrathin water layer was confined between the gel and organic solvent. The amide monomers were incorporated in gels while the aldehyde monomers were dissolved in organic solvent. After a period of time, an ultrathin film was formed between water/organic interface. Because of the slow release rate of monomers, the film was crystalline as confirmed by GIWAXS characterizations. Interestingly, this film was oriented, and their c-axis orientations were centered on the surface of COF film. While this method can be applied to other COF systems, it can also be expanded to include the preparation of MOF films as well.

Recently, Park et al developed a novel laminar assembled polymerization, by which wafer-scale monolayer 2D polymer (2DP) film is obtained.[22] The Wafer scale 2DP is produced at a pentane/water junction. In this home-built setup, solution with porphyrin-based monomers were injected between two phases by small nozzles. Once delivered to the interface, the monomers were spread and generated a laminar flow that confined within two phases. Subsequently, the polymerization occurred with the reagent present in water. During the growth, the pentane/water interface was steady, resulting in minimized distortion. The laminar flow was confined between the two phases, which led to wafer-scale monolayer. This method has potential to be extended to other systems to prepare large-area monolayer COF films.

### 3. Mechanical Properties

While much focus has centered on the performance of 2D COF films and membranes for practical applications such as high-pressure driven filtrations, usage in flexible devices, and membranes for batteries, all of these are directly related to the mechanical properties. Theoretical studies have suggested that 2D COF thin films possess outstanding weight-normalized mechanical properties.[41,89] Yet, experimental evaluations are still very limited. A thorough assessment of ultimate stress and Young's modulus data reveals that the mechanical performance of COF films is determined by their structures and compositions. For the former, increasing interlayer interactions such as $\pi$-$\pi$ or hydrogen bonding helps improve mechanical strength. Examples of the latter are hybrid materials such as COF-polymers or COF-cellulose fibers which outperform the corresponding pristine COFs in terms of flexibility and durability. The method of preparing thin films also plays an important role in shaping their mechanical robustness. Therefore, all of these factors should be taken into consideration while designing COF films with desirable mechanical performance.

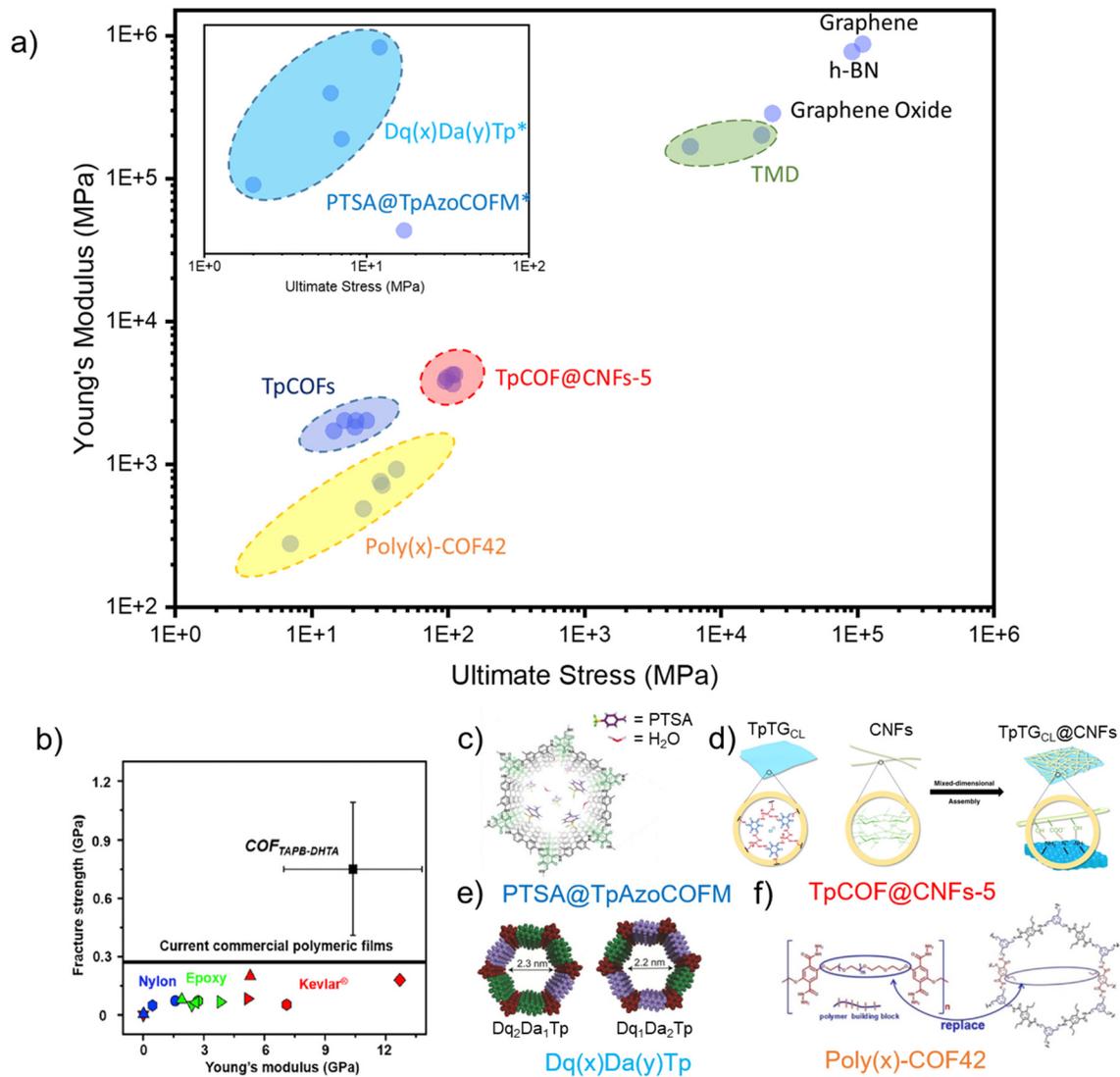

**Figure 6**: Mechanical properties of COFs compared with other functional materials. a) Young's modulus vs ultimate stress of reported 2D COF compared with values of TMD, graphene oxide, h-BN, and graphene; inserted graph: COF films/membranes with reported ultimate stress value only. b) mechanical properties of ultrathin COF$_{TAPB-DHTA}$ thin films compared with values of other polymeric films, c) structures of 2D COFs with reported mechanical properties including PTSA@TPAzoCOFM d) TpCPF@CNF-5 e) Dq(x)Da(y)Tp f) Poly(x)-COF42. b) Reproduced with permission from Ref [90], Copyright 2021, Elsevier, c) Reproduced with permission from Ref [91] Copyright 2018, WILEY-VCH Verlag GmbH & Co. KGaA, https://doi.org/10.1002/anie.201804753, d) Reproduced with permission from Ref [11] Copyright 2019, Springer Nature, e) Reproduced with permission from Ref [92] Copyright 2018, American Chemical Society, f) Reproduced with permission from Ref [93] Copyright 2019, American Chemical Society.

Integrating polymer chains into COF frameworks is an innovative approach to help boost the flexibility, processability and mechanical strength of a COF while maintaining unique advantages such as crystallinity and porosity. This method was utilized to construct a series of membranes (polyCOF) based on COF-42.[93] Specifically, the new linker DHT-400 was synthesized where 2,5-diethoxyterephthalohydrazide (DTH) was incorporated into a linear polymer with PEG-400 (Figure 6c). $Poly_x$COF-42 was then synthesized using the reaction between the linker mixture of DHT and x% of DHT-400 and 1,3,5-triformyl benzene (TB). Tensile tests indicate that both Young's modulus and ultimate stress increase with x when compared with those of pristine COF-41 (Figure 6a). Additionally, $poly_{3/6}$COF-42 has the highest value of Young's modulus (914 MPa) and ultimate stress (42 MPa), which were more than 2 orders of magnitude higher than the parent COF-42 membrane. Moreover, $poly_x$COF-42 also showed stronger mechanical strength as compared with other COF-membranes (Figure 6c). These results support the role of polymer chains in improving mechanical properties of COF membranes

Among those COF membranes with reported mechanical properties, TpCOF@CNFs-5 appeared with the most impressive mechanical properties (Figure 6a and 6c).[11] TpCOF@CNFs-5 was designed based on the mixed-dimensional assembly strategy between 2D COF nanosheets and 1D cellulose fibers (CNFs). Specifically, a series of β-ketoenamine COFs derived from 1,3,5-trisformylphoroglucinol (Tp) with different diamine linkers including $TpTG_{Cl}$ ($TG_{Cl}$ triaminoguanidinium chloride), TpPa-1 (Pa-1 p-phenylenediamine), TpBD (BD benzidine), TpHZ (HZ hydrazine hydrate), $TpBD(OH)_2$ ($BD(OH)_2$ 3,3'-dihydroxybenzidine) were chosen as 2D materials for this study. CNFs were oxidized by TEMPO to increase the density of carboxyl groups, thus strengthening the interaction between CNFs and COFs. Blending Tp COFs and oxidized CNFs in aqueous solution yields mixed-dimensional nanocomposites TpCOF@CNF-X where X corresponds to content of CNFs. Measuring tensile strength and Young's modulus of TpCOF@CNFs-X series revealed the much stronger mechanical strength of these hybrid materials as compared with the pristine TpCOF (Figure 6a). Particularly, $TpTG_{Cl}$@CNFs-5 membranes showed very high value of tensile strength (108.3 MPa) and Young's modulus (3.62 GPa). Thus, the combination of 2D COFs and 1D CNFs establishes robust interlamellar microporous network. Together with multiple interactions between COF and CNFs, this technique helps strengthen the mechanical properties of COF membranes.

Mechanical properties of 2D COF thin films can also be evaluated using in-situ technique such as AFM nano-indentation[88] or SEM mechanical testing.[90] The former was employed to measure Young's modulus of $COF_{TTA-DHTA}$ thin films. $COF_{TTA-DHTA}$ thin films were synthesized from 4,4',4''-(1,3,5-triazine-2,4,6-triyl)trianiline (TTA) and 2,5-dihydroxyterethaldehyde (DHTA) at the oil/water/hydrogel interfaces. Employing the superspreading technique, free-standing films with controllable thickness and good mechanical strength (Young's modulus value of 25.9 ± 0.6 GPa) were obtained.

Recently, a detailed study on mechanical properties of ultrathin COF films using *in situ* SEM testing technique were reported by Fang et. al.[90] In this work, TAPB-DHTA COF films were grown on sapphire substrates and meticulously transferred onto suspended sample stages of the SEM. Mechanical testing for fracture strength and Young's modulus of $COF_{TAPB-DHTA}$ films revealed values of 0.75 ± 0.34 GPa and 10.38 ± 3.42 GPa, respectively. These values show that the $COF_{TAPB-DHTA}$ films has a comparable mechanical strength with Kevlar thin films, as well as nylon- and epoxy-based polymeric materials (Figure 6b). Besides, *in situ* SEM mechanical testing of $COF_{TAPB-DHTA}$ films also provide valuable understanding of crack propagation, fracture toughness of the material, as well as its behavior toward fracture and pre-crack defects. Thus, this study provides an in-depth understanding of the mechanical robustness of 2D COF films and lays a foundation for their future applications.

## 4. Applications of Designer COF Thin Films

### 4.1 Filtration Applications

Because of their high porosity, tunable pore size, modular surface structure and well-defined pore structure, COFs are considered promising candidates for membrane separation for selective nanofiltration with excellent permeability. In this section, we discuss the gas separation and nanofiltration performance of COF-based membranes.

<u>Gas separation</u>

Crystalline porous materials, including zeolite, MOFs and COFs have long been considered in high performance gas separation membrane applications. However, progress of applying COF membranes has been limited because 1) preparation of continuous and crack free COF membranes is very challenging and 2) the pore size of COFs is relatively large (1 to 5 nm) compared to the kinetic diameter of gas molecules (0.2 to 0.5 nm). Nonetheless, significant efforts have been made to advance this important research direction. One way to address the second issue is to narrow the pore size. Caro et al. reported a continuous membrane based on azine-linked ACOF-1.[94] This COF has a pore size of 0.9 nm and the membrane exhibited an excellent separation factor of 86.3 in $CO_2/CH_4$ mixed gas separation with a favorable $CO_2$ permeance of about 9.9 x $10^{-9}$ mol $m^{-2}$ $s^{-1}$ $Pa^{-1}$. This high gas selectivity can be attributed to the intergrown grains in ACOF-1, in which large amounts of stacked pores were formed, narrowing pore aperture sizes. Another way is to enhance the affinity of gas molecules with pore walls. With chemical modification of the pore walls, the pore size would be decreased, and the functional groups would enhance the interaction with gas molecules. Jiang et al. first introduced active ethynyl groups on the pore walls of an imine-based COF.[95] Further modification enabled anchoring of functional groups within the pore wall to facilitate separation amongst gas species. The highest $CO_2$ uptake of 157 mg $g^{-1}$ at 273k, 1bar was achieved by optimizing the functional groups inhabiting the pore. The enhanced selectivity owing to functional groups grafting can be a useful strategy to enhance the gas selectivity within COF devices.

Another computational study based on few-layered COF membranes unveiled that, by changing the stacking order, the selectivity over $CO_2/N_2$ could change from non-selective to highly selective.[96] However, in most COFs, the most common stacking order is AA stacking, which is not able to provide the "gating-close" effect on the selective transport of gas molecules. In 2016, Qiu et al. reported a novel COF-MOF composite membrane.[97] ZIF-8 was directly grown on the surface of COF-300 membrane. At the interface of the COF and the MOF, the gaps between COF crystals were filed with amorphous MOF. This amorphous layer would seal the interface and showed high $H_2/CO_2$ selectivity ratio of 13.5 at room temperature, 1bar, exceeding the performances of both COF-300 membrane (6.0) and the ZIF-8 membrane (9.1). However, while an improvement, it was not as high as expected based on theoretical evidence. Later, Caro et al. prepared COF-COF membrane by successively growing ACOF-1 on the surface of COF-LZU1.[98] At the contact interface, interlaced pore structure was formed. According to the BET data, these pores had a diameter of 0.3-0.5 nm, much smaller than individual COFs (0.99 nm for ACOF-1 and 1.87 nm for COF-LZU1), showing excellent gas selectivity for both $H_2/CO_2$ (26.7), $H_2/N_2$ (88.7) and $H_2/CH_4$ (105.0) at room temperature. Because of the strong imine bonds, this membrane also exhibited long term stability. These results indicated that the formation of compact interfaces is the key to improve gas selectivity. However, while successful, both films were relatively thick, which limited their gas permeation. To enhance the interface property, more recently, Zhao et al. prepared a novel ultrathin membrane based on layer-by-layer assembly of oppositely charged ionic COF thin films (Figure 7a).[99]. Because of the strong interaction between these ultrathin films by electrostatic forces, this hybrid film exhibited reduced aperture size with a selectivity factor of 22.6 over $H_2/CO_2$ at 423K, surpassing most 2D material based membranes as well as the Robeson upper bound of conventional polymer membranes (Figure 7b). This film was ultrathin (41 nm) compared to previous works (around 1 µm), and showed a $H_2$ permeance of 2566 gas permeation units, indicating the promising potential of ultrathin COF films in gas separation area which is comparable to top performing materials such as $Zn_2(bim)_4$. (Fig. 7b).

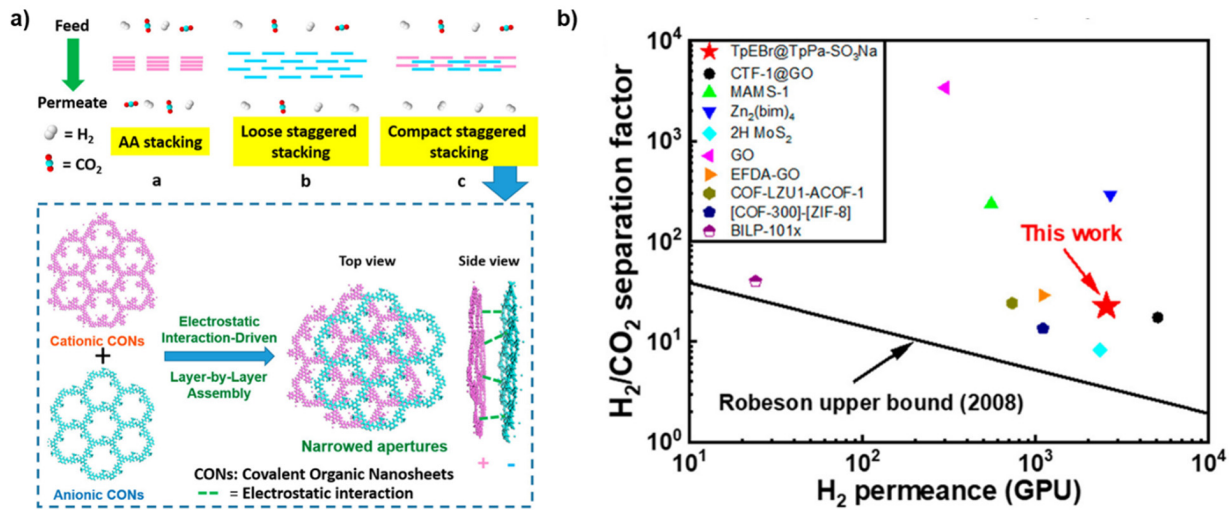

**Figure 7**: Performance of COF-based gas separation devices. a) Schematic illustration of compact stacking of two ionic COF b) Comparison of $H_2/CO_2$ separation performances with other membranes. Reproduced with permission from Ref [99] Copyright 2020, American Chemical Society.

Nanofiltration

Crystalline nanoporous materials, like zeolites and MOFs, exhibit superior filtration property over tradition polymeric membranes.[100,101] However, these materials are relatively unstable in aqueous condition. Compared to zeolites and MOFs, the covalent bonds in COFs promotetheir stability in both aqueous and organic solvents. Thus, COFs are increasingly considered to be effective and desirable materials for nanofiltration. As previously described one of the major challenges is to prepare large area, crack free membranes. In 2017, Banerjee et al. demonstrated that large area, crack free COF membranes could be prepared by liquid-liquid interfacial polymerization.[66] These freestanding membranes were then transferred to supporting substrates for further nanofiltration tests. The result showed that these membranes demonstrated remarkable solvent-permeance. Take Tp-Bpy for example, the acetonitrile permeance reached 339 L $m^{-2}$ $h^{-1}$ $bar^{-1}$. All investigated COF membranes exhibited high rejection (over 90%) rate to organic dye molecules, and these COF membranes relied on the natural pore size effect to exclude molecules. Through framework modifications, Ma et al. prepared a cationic COF membrane using similar method (Figure 8a).[102] This cationic framework incorporated ionic units, thus the mass transport in nanochannels was not only controlled by pore size but also by the charge. The cationic COF membrane exhibited especially high rejection rates for anionic dyes (>98%) (Figure 8b), as well as high permeance for water (546 L m-2 h-1 bar-1).

Additionally, COF thin films have also be applied in water desalination. However, the COF pore size is larger than hydrated salt ions, resulting in the low rejection rate of salt ions.

Lai et al. reported the preparation of ultrathin COF film on porous polysulfone support.[75] The pedant groups on the pore wall reduced the pore size to 1.5 nm. This membrane exhibited rejection rate of 64% and 71% for NaCl, Mg$_S$O$_4$, respectively. To improve the rejection rate to salt ions, Huang et al. modified COF films with carboxyl groups.[103] After the modification, this COF film showed excellent ion rejection rate (>90% for most salt ions) with a water flux of 0.5 L m$^{-2}$ h$^{-1}$ bar$^{-1}$ (Figure 8c). The introduction of carboxyl groups can reduce the pore size and the negative charge on carboxyl groups can effectively limit salt ions passing through. These results indicated potential water desalination performance improvement of COF thin films with more carefully designed functionalization schemes.

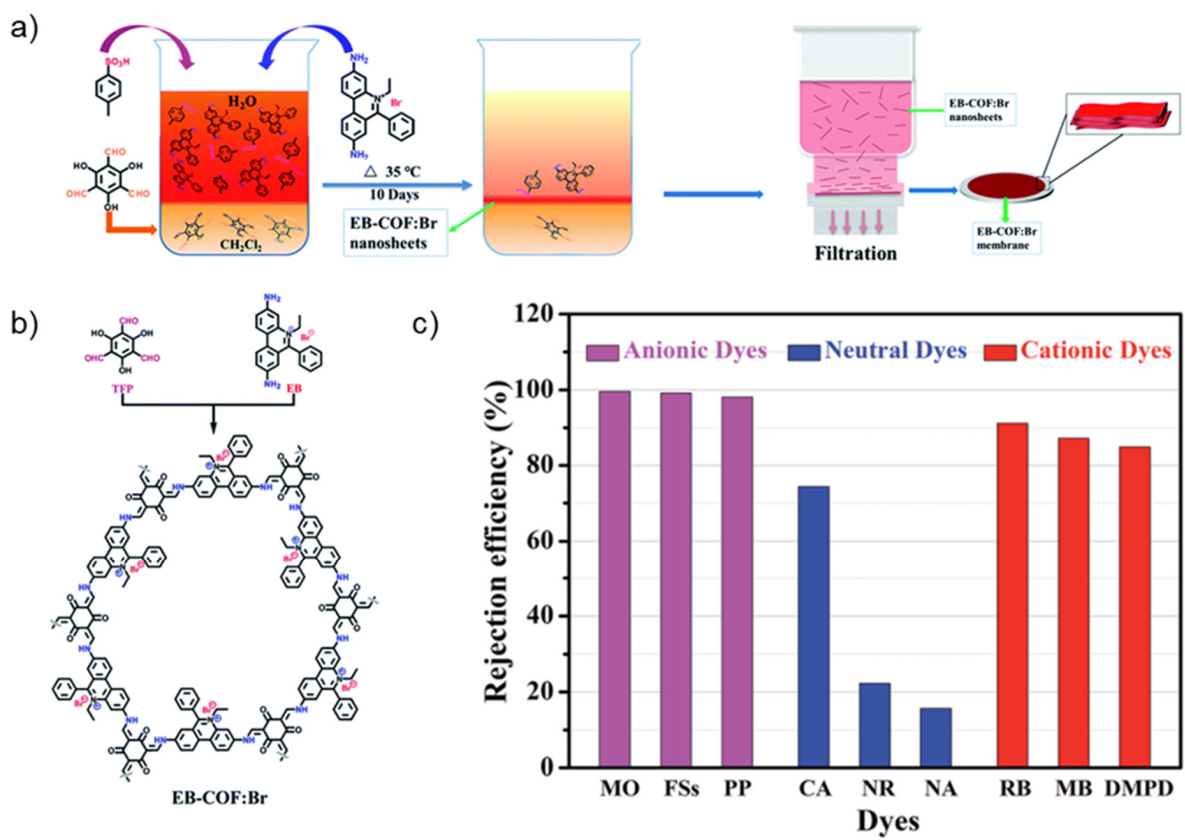

**Figure 8**: Nanofiltration applications of COFs. a) Schematic illustration of the COF film preparation method, b) structure of cationic COF, and c) the rejection efficiency of the COF membrane for different dyes. Reproduced with permission from Ref [102]. Copyright 2018, The Royal Society of Chemistry.

### 4.2 Electronic Properties of COF Thin Films

Attaining high device performance from scalably-synthesized polymers is challenging, as the materials are characterized by at least some degree of disorder embodied as charge trapping defects and localized electronic states in disordered regions. Covalent organic frameworks are conjugated polymer chain assemblages with organized paths for charge carriers within and/or between layers with ever increasing values for charge mobility as new architectures with charge transport enhancing topologies are uncovered. In addition to semiconducting properties exceeding those of amorphous and even polycrystalline silicon, and possessing advantages over semiconducting polymers such as environmental stability (especially compared to n-type semiconducting polymer materials), additional benefits of COFs include ease of processing (especially in 2D form), mechanical flexibility,[93] controlled pore size,[104] extreme surface areas, and environmental stability (up to >300 °C),[105] COFs are attractive for diverse applications including low-cost, flexible, and wearable devices.

Composition and ordering of COF building blocks to engineer charge transport

In polymers, the covalent nature of bonding and significant electronic localization give rise to relatively large electronic band gaps, however the unique architecture of COF materials promotes charge delocalization for increased carrier mobility. Therefore, in COFs, electronic properties are dictated by the individual building blocks, and the way in which they are arranged. In general, the semiconducting properties of conjugated polymers depend on the interaction between half-filled orbitals on adjacent carbon atoms along the polymer backbone. Distortion, particularly that associated with formation of a one-dimensional network from polymerization leads to pairing of electrons, with chains of alternating single (σ) and double (π) bonds, hence the use of the term π-conjugated polymers.[106,107] While for isolated chains, charge transport is one-dimensional along the backbone, in molecular assemblies, charge transfer between chains may also contribute to increased mobility.

The nature of the charge carrier is strongly dependent upon the selection of COF building blocks.[108] COFs can demonstrate any possible semiconducting behaviors (Figure 9), depending on selected repeat units. P-type semiconducting COFs generally result from extended π-conjugated systems.[37] The first example of n-type conduction employed electron accepting benzothiadiazole (BDTA) at the edges of a nickel-pthalocyanine COF.[109] For ambipolar conduction, where electrons and holes function as charge carriers, COFs can be assembled from alternately linked donor and acceptor π-arrays.[110] Another option is a supramolecular approach with spatially confined electron acceptors within the open channels of electron-donating frameworks such as skudderites or buckyballs.[111]

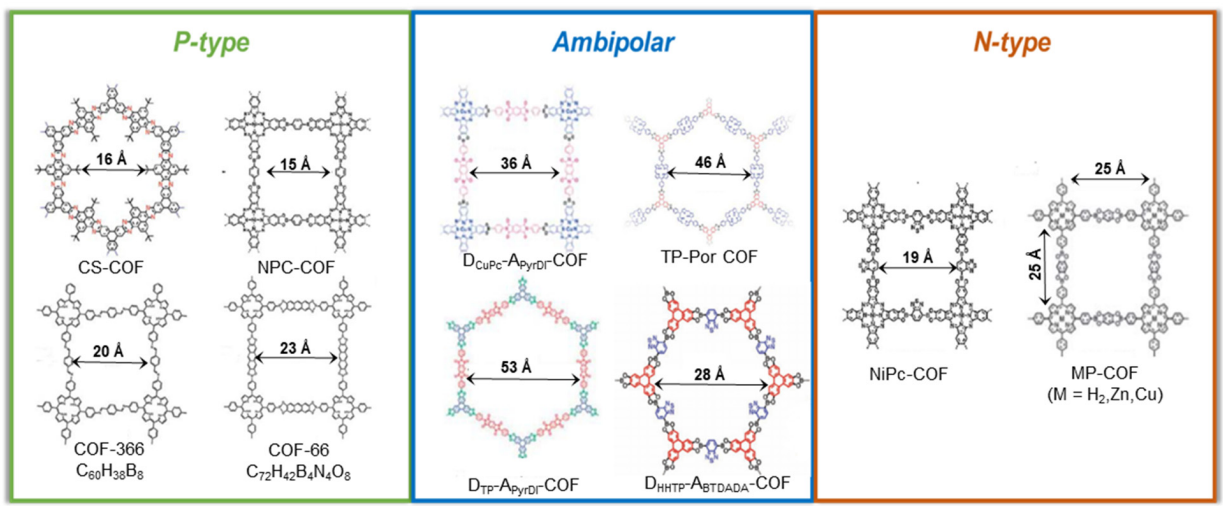

**Figure 9**: Example building blocks of p-type, n-type, and ambipolar COFs. Reproduced with permission from Ref [108], Copyright 2021, Elsevier.

Charge transfer within organic semiconducting materials depends upon both electron-electron interactions and electron-phonon interactions.[112] These phenomena can be characterized by two quantities: (1) the transfer integral, which provides insight on the degree of π-π bonding between neighboring molecules and (2) the reorganization energy which is an indicator of the energy required for molecules to go from an initial state prior to charge transfer to the final state after the process is complete.[113] These quantities are related to the extent and nature of molecular packing within the assembly, the intermolecular interaction strength, and the chemical structure of π-conjugated organic semiconductors.[56,114] In general, greater charge mobility is realized with π-orbital overlap between ordered molecules within an array,[115] although additional factors such as impurities, defects, and charge carrier density can all affect electronic properties.

Fundamental tools to adjust π-π overlap within individual building blocks include conjugation length, aromaticity, and heteroatom replacement to delocalize electrons to reduce the electronic band gap. For example, an experimental study of linear acenes from anthracene to hexacene revealed incrementally increasing charge mobility with each addition of a ring from 0.15 to 1.5 $cm^2$ $V^{-1}s^{-1}$ with a reduction in band gap from 5.2 to 4.4 eV.[116] Applying these tools on COF has also been explored effectively via simulations. For example, chain lengths for HHTP-DPB-COF, COF-1, and CTF-1 were all varied systematically with accompanying reductions in the electronic band gap for each COF(up to 30%) due to distribution of the electron density in the valence band maximum.[117]

While the intrinsic properties of COF building blocks affect electronic properties, the architecture of the layers also play a significant role. Examples of building blocks with poor conductivity morphing onto strong conductors are ZnPc-pz and CuPc-pz, which both demonstrate localization of electron density with flat bands and low in-plane charge transport. The situation changes, however, for stacked COF layers, where appreciable dispersion of the band structure of the COF is observed. However, for a particular stacking configuration, the band gap of crystalline ZnPc-pz is narrowed from 1.2 to ~0.6 eV in COF form with hole mobility >5 cm$^2$ V$^{-1}$ s$^{-1}$.[105]

Linear polymers demonstrating large intermolecular electronic conduction are rare, however selection of building blocks with attractive properties for integration into two-dimensional COF sheets, such as porphyrin units, possessing close intermolecular π-π distances. Blocking linkages block transport between sheets, however, with this approach, where all the atoms within the structure were completely superimposed upon those of an adjacent sheet, movement of charge carriers from one sheet to another via eclipsed integration of π-electronic components into a well-defined 2D layered frame work was enabled for the formation of two new COFs (COF-366 and -66) yielding extreme high hole carrier mobility values of 8.1 and 3.0 cm$^2$ V$^{-1}$ s$^{1}$, respectively.[118]

Tailored conducting and dielectric properties

Incorporation of dopants, adding side chains, mechanical strain, and integration of COFs into heterostructures are all means by which the electronic properties of COFs can be adjusted. Effective electronic doping is facilitated by key structural features of COF films (Figure 10). Alignment of electrically conducting columns coupled with open nanoscale channels for binding of atoms and molecules[119] allows integration of dopants selected for strong charge transfer interactions as reported for tetrathiafulvalene (TTF)-based thin film COFs.[9] TTF is a strong electron donor forming crystals that readily exchange charge with strong electron acceptors such as iodine or tetracyanoquinodimethane (TNCQ). Electrical conductivities of up to 0.28 S m$^{-1}$ were measured for TTF COFs exposed to iodine vapor. The increase from doping was reversible as the samples were removed from an iodine rich environment. The conductivity changes are correlated with diagnostic signatures in UV-vis-NIR and EPR spectra, pointing to effective radical delocalization within mixed-valence TTF stacks. Optimization of channels within similar COF materials via different chemical linkages or spacers is a likely path for increased conductivity. A similar vapor phase iodine scheme was used in porphyrin-based COF materials increasing electrical conductivity over three orders of magnitude.[120]

Correlations of strain to electronic band structure for low dimensional materials such as 2D graphene,[121] MoS$_2$[122] are well-known, but only computational studies on effects of strain on COFs currently exist in the literature. DFT simulations of the relaxed structure of COF-1 under uniform strain by scaling of the lattice constants showed reduced band gaps of up to 30% (from 3.5 to 2.5 eV) predicted with application of elastic tensile strain and increases under compressive strains less than 2%.[117]

While controllable porosity is an attractive feature of COFs and materials such as porous alumina demonstrate extreme low dielectric constant (k), the values of k for porous COFs demonstrates sensitivity to water absorption in typical ambient environments with 40-60% relative humidity.[123] Manipulation of COF building blocks allows for stable tuning of dielectric properties, even in the presence of humidity. Record low dielectric constants are reported for COFs prepared via interfacial polymerization covalently linking tris(4-aminophenyl)-benzene (TAPB) and terephthalaldehyde with alkoxy groups of different lengths (TPOCx) with x varied incrementally from 1-5.[84] Increasing the length of the alkoxy chains resulted in a decrease of k from 1.55 to 1.19, stable in the presence of humidity and over 1000 mechanical bending cycles. Young's modulus values of the films correlated to dielectric properties and the different packing compactness of COF particles in the films. Water contact angles of the TAPB-TPOCx-COF films show that the hydrophobicity also increased with increasing the alkoxy chain length, promoting stability. Figure 2c shows values of these materials in contrast to values reported for other inorganic and organic ultralow-k dielectric materials such as FPTTPI and Teflon.[124]

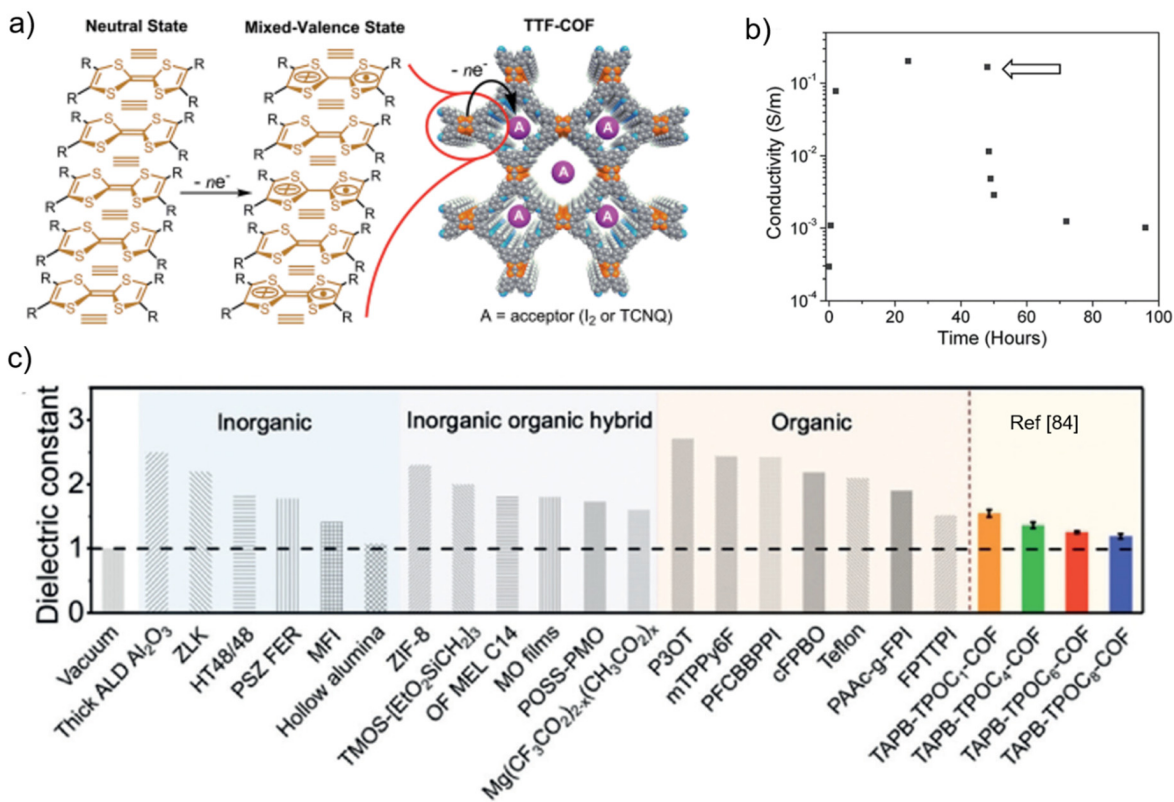

**Figure 10**: Electronic conductive and dielectric properties of COFs. a) TTF-COF electronic states with b) tunable conductive properties and c) the dielectric constants of familiar materials (left in gray) in comparison to state of the art dielectric COF films.[84]

Reproduced with permission from Ref [9]. Copyright 2014, The Royal Society of Chemistry, reproduced with permission from Ref [84].

### 4.3 Electrochemical applications

Because of their highly tunable structures, COFs exhibit a broad range of unique physical and chemical properties. By designing the node or linkage building blocks, COFs can achieve tailored properties with applications in diverse technologically significant areas. Electrochemical applications of COF films, including electrochemical catalysis, batteries, and capacitors are important areas of focus in current research. Due to the higher degree of accessibility of functional groups in thin films, COF films usually exhibit better performance in these electrochemical applications, compared to their bulk counterparts.

Electrochemical catalysts

In 2015, Yaghi et al had developed a cobalt containing COF as electrocatalyst for $CO_2$ reduction to CO in aqueous media, in which cobalt centers served as electrochemical active sites (Figure 11).[62] The COF based catalyst exhibited both high Faradaic efficiency (90%) and turnover numbers. However, because of the low solubility and limited $CO_2$ diffusion in bulk COFs, only a small portion of active sites (cobalt centers) were utilized. Subsequently they prepared oriented thin films on glassy carbon and fluorine-doped tin oxide (FTO). The oriented thin film provided more accessible active sites and showed increased catalytic efficiency. As a follow-up to the oriented thin film work, a series of oriented cobalt-containing COF thin films were prepared with modified electron structures.[60] Their results show that COFs grown as oriented thin films significantly improved their catalytic performance over bulk samples and optimization of activity and selectivity through facile modification of side groups within the film,.

Capacitors

Another important application of COFs in the electrochemistry field is as a capacitor. To date, most of the COF-based capacitors were based on anthraquinone (AAQ) structures in which they are assembled into periodic networks within the 2D structure. In 2015, Dichtel et al. systematically studied the characteristics of AAQ-based COF capacitors, preparing both powders and oriented thin films.[125] Cyclic voltammetry experiments indicated that more AAQ groups in oriented COF films were electroactive compared to bulk powders, in which only 3% were accessible. The galvanostatic charge-discharge cycles revealed that the charge storage capabilities would be dramatically improved with the oriented COF films. Later it was seen that freestanding AAQ-based COF films were prepared by a slurry-casting method, the redox-active COF film displayed extreme stability even in concentrated sulfuric acid ($H_2SO_4$).[126] Moreover, this freestanding film exhibited extraordinarily high areal capacitance of 1600 mF $cm^{-2}$. These results demonstrated that AAQ-based COF thin films have great potential in applications as capacitors in devices.

Lithium batteries

Tailored structures of COFs have shown proven potential for organic electrodes in lithium batteries. With the introduction of redox-active groups into the skeleton, COFs provide the scaffolding for abundant litium storage sites.[127–129] Until recently, most of the promising demonstrations of COF-based electrodes were in the powder form and not in thin films. In 2017, Wang et al. found that, when exfoliated down to thin films, COFs provided shorter lithium transport pathway as well as more accessible active sites. Thus compared to bulk powders, COF thin films exhibited improved battery performance, including capacity, cycling stability and rate behavior.[130] Based on this work, Feng et al showed direct grown highly stable, crystalline 2D polyarylimide thin films on the surface of carbon nanotubes.[131] This hybrid structure featured abundant redox-active naphthalene diimide units, highly accessible pore structures, excellent structural stability, and fast lithium diffusion. As the result, the rate capacity and cycling stability far surpassed state-of-the-art polyimide electrodes.

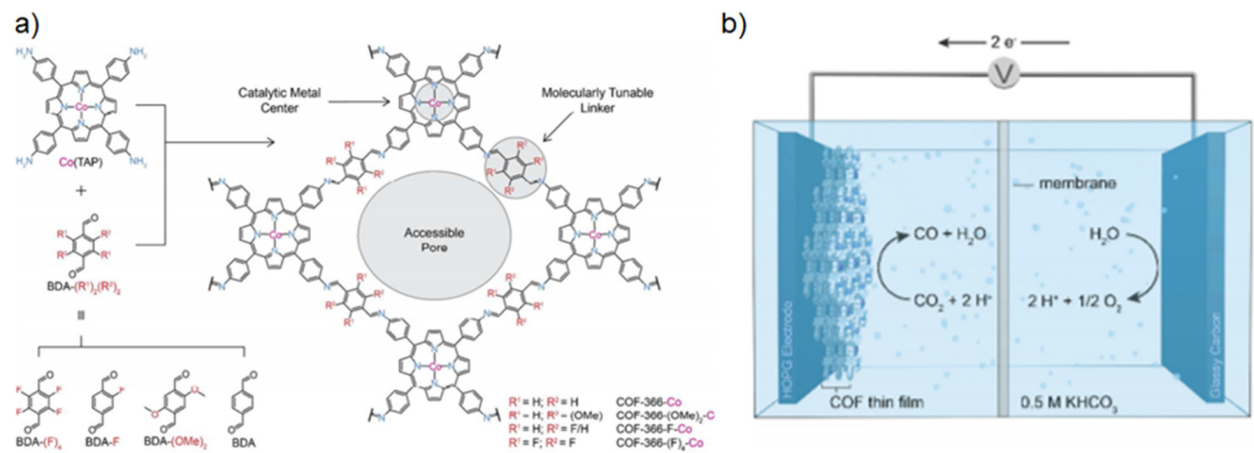

**Figure 11**: Electrochemical catalytic properties of COFs. a) Schematic illustration of cobalt-containing COF structures. b) Schematic illustration of electrolysis cell for electrochemical $CO_2$ reduction. Reproduced with permission from Ref [60]. Copyright 2018, American Chemical Society.

### 4.4 Optical Properties and Optoelectronic Devices

Absorption and Emission

The highly conjugated aromatic regions typical for COFs promotes modular incorporation of interesting chromophores for useful interactions of the framework with incident light.[132,133] Coupling the primary structure with the strong π-π interactions from the available conjugated π-systems located throughout the primary aromatic framework [14], several high order donor-acceptor interactions can be tuned to interact with a wide range of intensities of incident photons, even with few layers of these frameworks in thin films (Figure 12a).[104,134] The intense π-π stacking events cause the rapid quenching due to thermal decay of the photoexcitation in photocarrier systems, with decreasing fluorescence as the number of layers increases. However, Jiang and Coworkers later showed that injection of Ammonia resulted in a framework layer disaggregation and subsequent increase in ammonia concentration-driven COF fluorescence.[134] Post processing and activation of COFs to induce desired geometries to enhance or decrease emission was demonstrated by Wang and coworkers. Upon activation of BZL COF resulted in decreased interlayer distance, giving rise to AIE changes in phosphorescence (Figure 12b-c).[135] Kuc et al showed that a staggered interaction allowed for increased band bending.[136] These aggregate-caused quenching (ACQ) and aggregate induced emission (AIE) phenomenon are shown to significantly increase and decrease the expressed signal in both fluorescence and phosphorescence. COF-containing thin films therefore demonstrate a broad range of tunability based on the building blocks selected for the primary structure.[41]

Chemical post-synthetic modification of the organic framework is focused on the addition or removal of protons from the Brønsted-Lowry Acid and Bases that join and/or line the linkers and vertices throughout the framework. In the case of work from Jiang and coworkers, the addition of fluoride ions act as proton scavengers, this leads to the subsequent deprotonation of the hydrazone linkage.[137] This transformation causes a disruption of the conjugated aromatic framework and transfer of electrons consequently increasing the luminescence within the framework (Figure 13a). Due to the reticular nature of COF synthesis, the geometry of the primary structure of the COF is locked upon formation.[38] However, the pore structure can be tuned to favorable or unfavorable pore geometries based upon photoexcitation-induced isomerization.[138] A marriage of these influences is evidenced by the work of Li et. al., the coupled photoinduced tautomerization is a combination of the photoexcitation and the presence of the solvent to change the activity within the system.[46] In the case of Yan and coworkers, water was used to influence the dual-emission system through both intramolecular charge transfer (ICT) and excited-state intramolecular proton transfer (ESIPT).[139] These post-synthetic shifts upon photoexcitation are interesting ways to influence not only the optoelectronics but pore shape and therefore filtering and sensing applications.

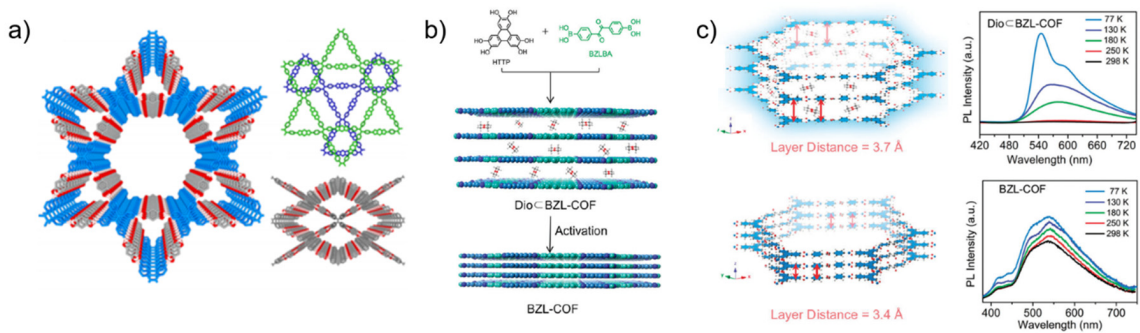

**Figure 12**: Schematic of stacking arrangements of COF in relation to optical emission output. a) COF design to incorporate stacking pattern (AA, AB, and single pore rhombic AA) to maximize aggregation-induced emission (AIE) and prevent emission suppression inherent in many thin film COF thereby increasing quantum yields, b) activation of COF via decreasing the interlayer distance affecting the AIE and ACQ c) Phosphorescence emissions increasing through a reduction of interlayer distance. a) Reproduced with permission from Ref [134], Copyright 2016, American Chemical Society, b-c) Reproduced with permission from Ref [135], Copyright 2018, Royal Society of Chemistry.

Catalysis of chemical reaction via solid phase substrate has proven to be an important aspect in green chemistry increasing the efficiency and durability of the catalyst which often leads to recyclable materials.[27,49,140,141] Interestingly it is important to note the increasing attention paid to photocatalyst COFs[17,142] The incorporation of interesting chromophoric modalities in the framework allows for the tunable absorption of photons, these photons in turn are utilized to accomplish chemical transformations.[46,143–145], such as synthesis of Lp-pi-COF thin films carried out with photon harvesting of the precursors.[76] Commonly, photoisomerization is an important intrinsically energy-driven phenomenon, where the photon absorption and emission can be coupled to drive an isomerization event. Bhadra et. al. and coworkers showed the absorption, intersystem crossing and energy transfer causing isomerization of stilbene and stilbene-like derivatives from the low energy *trans*- configuration to the higher energy *cis*- isomer through the formation of a double radical stilbene species. This double radical mechanism reaction is inhibited in the presence of the radical scavenger TEMPO (4 equiv), indicating a quench after the intersystem crossing event.[146]

Isomerization doesn't necessarily only occur within the substrates, there is also potential for photoisomeric events to take place in the framework itself. While the addition of water or acid have been shown to cause chemically induced isomerization of the framework (Figure 13e-f),[46] photoisomerization can occur as well upon the absorption of photons (Figure 14a).[147] Photon absorption driven isomerization in COFs has been shown to occur, often to the detriment of the framework itself, such as with the incorporation of commonly photoisomerized targets such as aza-benzene systems.[148] Upon photoexcitation and isomerization there is a break in the framework lattice, while Lei and coworkers showed that upon annealing the framework is regenerated resulting in a self-healing optical sensor. Further, Lei and coworkers have shown that upon introduction of certain pore inhabiting species, an energetically favored bond swapping will occur and subsequent isomerization upon photoexcitation.[138] One of the hallmarks of optical properties is color, changes in wavelength of emitted or absorbed light are especially useful and important at the device level. Changes such as observed by Yan and coworkers,[139] where the emission of the COF have been altered due to the presence or absence of water due to a water induced tautomerization in the scaffold. Yan further demonstrated that their COF was able to act as a ratiometric sensor of water within the system (Figure 13b-d). Jiang and coworkers showed that by rational design changes and solvent-mediated interactions, changes in the emissions of sp$^2$ linked tetragonal highly chromophoric frameworks could be realized (Figure 13g-h).[149]

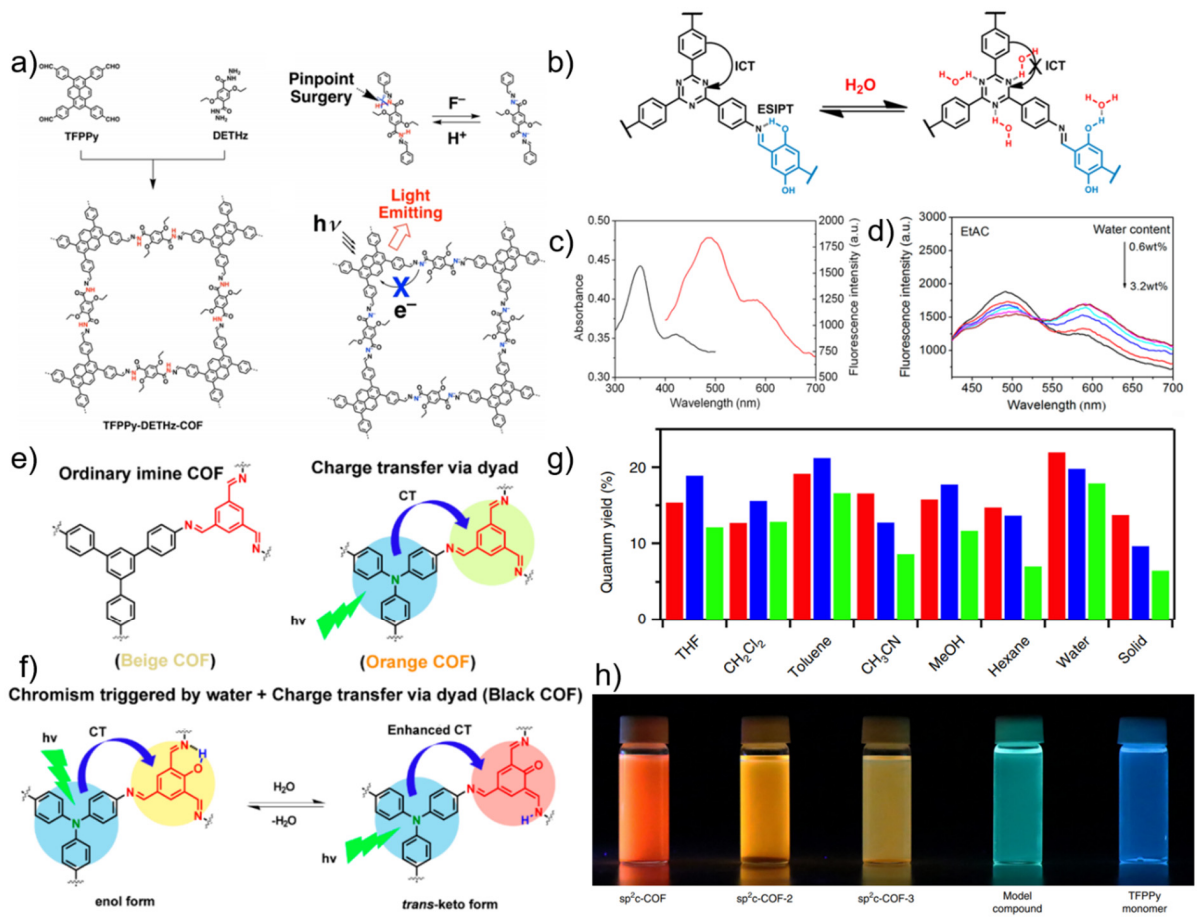

**Figure 13**: Examples of modifications for tailorable optical emissions of COF thin films. a) Chemical modification of COF pore to enable light emission, b) mechanism by which water prevents excited-state intramolecular photon transfer (ESIPT) due to isomerization in TzDa-COF, c) Normal UV-Vis absorption (black) and fluorescence (Red) in TzDa-COF d) Fluorescence spectra of TzDa-COF in EtAc with varying water content, e) vertex substitution resulting in color change and charge transfer upon photoexcitation, f) photoexcitation and water induced tautomerization leading to enhanced charge transfer, g) Solvent-mediated alteration of the quantum yield h) linker-mediated alterations of emission for C=C linked tetragonal frameworks. a) Reproduced with permission from Ref [137], Copyright 2018, American Chemical Society, b-d) Reproduced with permission from Ref [139], Copyright 2017, American Chemical Society, e-f) Reproduced with permission from Ref [46], Copyright 2018, American Chemical Society, g-h) Reproduced with permission from Ref [149], Copyright 2018, Springer Nature.

Optical Device Constructs

Device level application and implementation of COF thin films is an important step in the utilization of these fascinating optically responsive materials. The optical property inherent in these materials in concert with the ease of synthesis and low processability lead to simple efficient thin films with a wide area of optoelectronic properties. These include the facile incorporation into existing device schema to yield often highly improved performance.

Sequestering chemicals as a method of detection or purification leads to a change in the optical properties of the COF.[150] These interactions, whether through chemisorption or electronic interactions, effectively result in electronic doping of the COF and can either enhance or inhibit the photointeractions.[151] An innovative approach was employed by Lei et.al with the use of a sacrificial layer to trap analyates within a device upon photostimulation.[138] Another important feature of these versatile materials is their ability to act as chemical catalysts upon photoexcitation. While several chemical reactions have been studied, perhaps the most interesting is their incorporation on the device level allowing for facile conversion of atmospheric carbon. Zhu et al showed photocatalysis of carbon dioxide to methanol with high efficiency and easily obtained materials (Figure 14b).[152] Innovation incorporating COF thin films for the conversion of carbon sources to more benign or useful forms and is an exciting and important advancement for the green chemistry community. Photoswitching was demonstrated via incorporation of $COF_{ETBC-TAPT}$ by Lu et. Al, into a graphene-based photodetector with impressive response times (Figure 14d),[59] whereas Choi and coworkers showed Lp-pi-COF on $SiO_2$ as both a photodetecting and humidity sensing device.[76] These graphene-based devices showed superior performance when considering the sensitivity to response time of other known model semiconductors and increasing the utility of COFs incorporated into flexible electronic devices. Importantly to changes in resistance is change in color of thin films, these thin films are engineered to have tunable HUMO-LUMO interactions upon exposure to certain chemicals as shown by Lu, Loh and coworkers[46] (Figure 14c). Whereas Banerjee and coworkers showed quenching or enhancement of emission spectra for their TpBDH- and TfpBDH-based COF nanosheets due to the interactions with various analytes.[40]

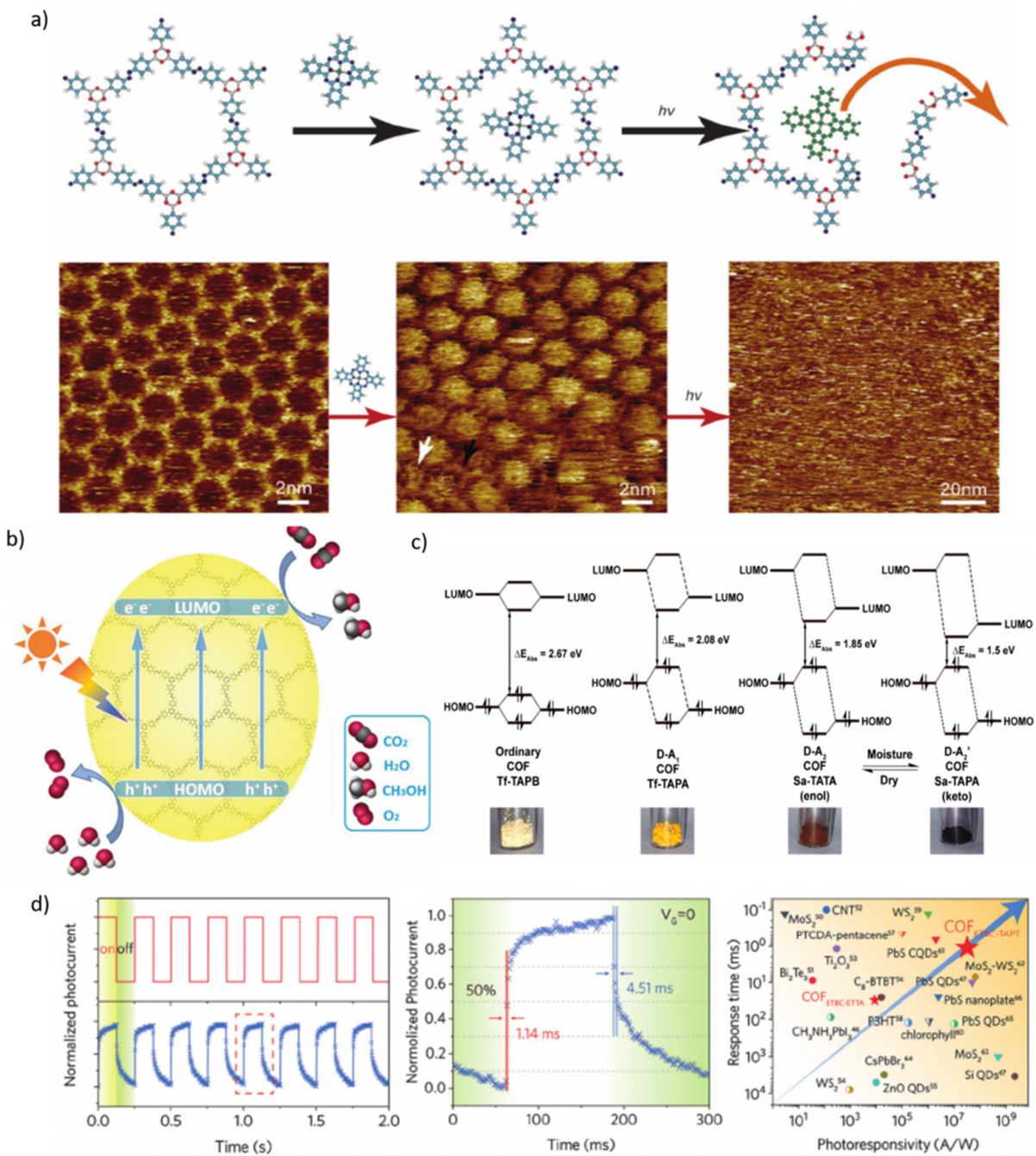

**Figure 14** Example optical device applications using thin film COFs. a) Photoisomerization and further bond swapping with analyte leading to deformation of pores and eventual breakdown of device, b) Photoinduced electron promotion causing $CO_2$ bound to the device to convert to methanol, c) Optical changes based on chemically induced tautomerization, d) Device photoswitching under alternating dark and light conditions and comparison to other photodetector systems. a) Reproduced with permission from Ref [138] Copyright 2016, WILEY-VCH Verlag GmbH & Co. KGaA,

https://doi.org/10.1002/chem.201601199, b) Reproduced with permission from Ref [152], Copyright 2018, Elsevier, c) Reproduced with permission from Ref [46], Copyright 2018, American Chemical Society, d) Reproduced with permission from Ref [91] Copyright 2020, WILEY-VCH Verlag GmbH & Co. KGaA, https://doi.org/10.1002/adma.201907242.

### 4.5 Other properties and niche applications

Many areas of COF thin film work are currently being focused on technologically significant issues and have yet to be fully realized and explored.[10,153–155] Several magnetic COF thin films are simple incorporation of ferromagnetic and non-ferromagnetic metal ions into the pore either by passive or active coordination to the framework itself. These allow for both templated COF synthesis or post-synthetic introduction of the magnetic property into the thin film.[156–159] However, Jiang and coworkers first saw that 2D sp$^2$ linked carbon polymers could be doped with iodine giving rise to ferromagnetic active frameworks.[160] Later work by Wang and Yu, showed that with purposeful design and matching of several dopants that stabile metal-free magnetic thin films are possible, thereby paving the way towards organic spintronic device applications(Figure 15a).[161]

Incorporation of interesting pharmaceuticals within or upon the framework are interesting applications, utilizing the framework for a time delay release in biorthogonal or nontoxic covalent networks in the treatment of chronic or acute conditions safely.[30,162,163] These frameworks show promising favorable interactions and biocompatibility which enhance their integration with other biological applications. Banerjee and coworkers showcase the TpASH series of drug delivery COFs for their facile synthesis and simple post-synthetic modifications.(Figure 15b).[164] Further work by Akyuz, showed that carbo-platin drugs can be stored in COF surfaces to prevent deleterious biological affects by slow release kinetics in the treatment of certain cancer cell lines.[29]

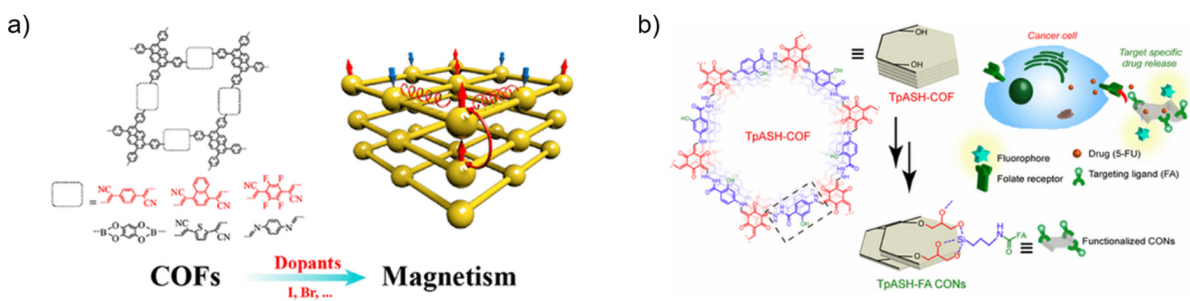

**Figure 15**: Alternative applications of COFs for magnetics and drug delivery. a) Construction of metal-free COFs with various electron donors, b) Schematic of TpASH-based COF thin films in the delivery of 5-Flurouracil. a) Reproduced with permission from Ref [161], Copyright 2020, American Chemical Society, b) Reproduced with permission from Ref [164], Copyright 2017, American Chemical Society.

## 5. Integration of COFs thin films with other materials

Electronic Heterostructures

A recurring theme when considering electronic properties of COFs is the role of periodic p-orbitals and their interactions with adjacent layers. Investigating the effects of mating COFs to substrates with complimentary electronic periodicity has inspired device fabrication for exploration of COF film-substrate interactions.[54] Graphene is a substrate with such an electronic arrangement in addition to numerous useful attributes as an electrode material.[36,165] Figure 16a shows a schematic of a COF-based vertical field effect transistor (VFET) with a single layer graphene source electrode.[56] The 1,3,6,8-tetrakis(p- formylphenyl)pyrene (TFP$_y$) and p-phenylenediamine (PPDA) COF material was ambipolar,[166] with significant hole mobility with gate voltages in the range of -10 to 2V and an on-off ratio of $10^6$ (Figure 16b). Electrons were less mobile than holes in the TFPy-PPDA COF over the range 1-10V due to the higher hole mobility in the COF p-columns and also higher collection efficiency for the higher work function gold electrodes comprising the device.

Application of COF films to versatile substrates such as graphene broadens application space to capitalize upon the mechanical flexibility, high electron mobility and large surface area. Vapor phase sensors are an ideal application where flexibility, conformality, and small area are all desirable attributes met by COF films. In one demonstration of a photocurrent-based vapor sensor, gas molecules interacting with COF$_{ETBC-TAPT}$, transfer electrons and change the charge carrier distribution within the heterostructure, affecting the photoelectric characteristics of the device. The COFETBC–TAPT in particular contained aldehyde groups in the pores, resulting in adsorption of $NH_3$ and other polar gas molecules and influence the photoelectric performance of the device with a dependence on the strength of the interaction allowing identification of particular gases based on the magnitude of the photocurrent response.

Finally, the 2D nature of COFs coupled with the unique electronic properties highlighted here invites their use in vertical heterostructures of 2D materials. Demonstration of wafer-scale COF integration in a sequence COF-2D transition metal dichalcogenide heterostructures for tunable capacitors highlight the potential for versatile electronics (as shown in Figure 16c-f).[22] Recent work has also demonstrated that 2D polymer and TMD heterostructures using a TIIP COF and monolayer $MoS_2$ moiety exhibited electronic coupling between the inorganic and organic components.[19] This important advance was followed up with further work by Hopkins et. al. which showed promising results when integrating of 2D COFs into heterostructures as thermally conductive, low-*k* dielectrics. These results reiterate the value of 2D inorganic/COF heterostructures as potential targets for use in integrated circuits due to their inherent low density, low-*k* dielectric and high thermal conductivity properties.[20]

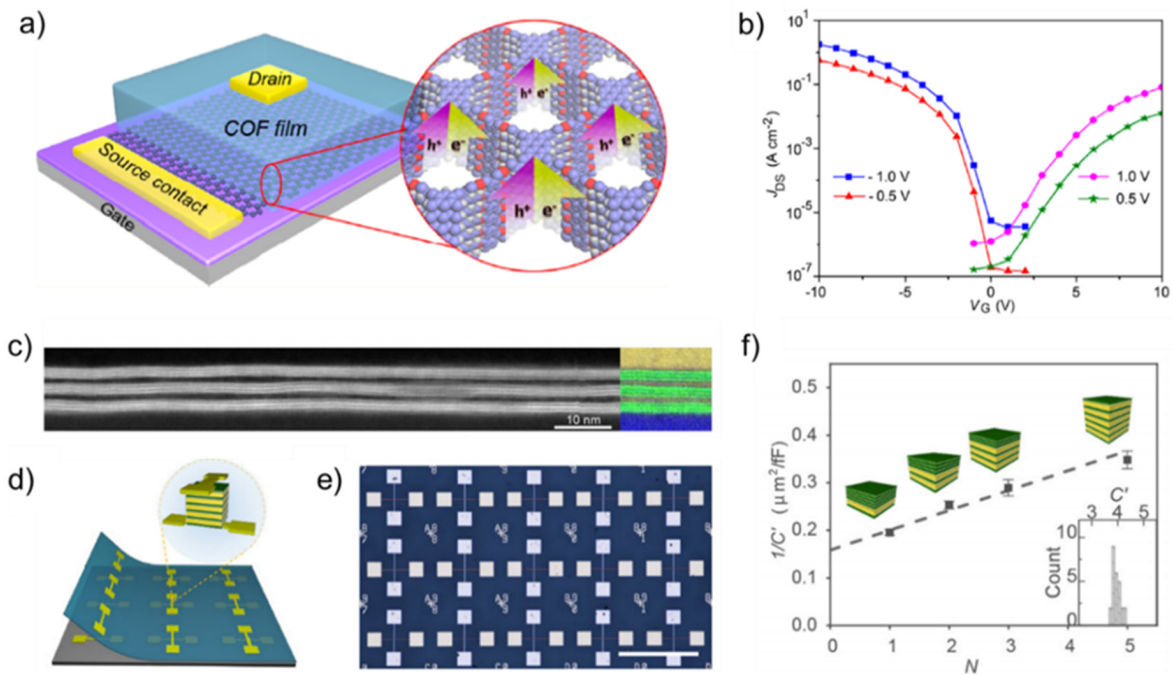

**Figure 16:** Device level design from semiconducting COF a) and b) integration of 2D COFs with other 2D materials for electronic devices. c) TEM image 2D COF-MoS$_2$ superlattice shown in cross section where the green false color depicts tri-layer MoS$_2$ between COF monolayers. d) schematic and e) micrograph of a wafer-scale heterostructure device array f) capacitance normalized by device area for heterostructures of with the number of repeat units. a-b) Reproduced with permission from Ref [56]. Copyright 2017, American Chemical Society. c-f) Reproduced with permission from Ref [22], Copyright 2019, Springer Nature.

Multifunctional Coatings

Chemically stable COFs have also been applied as coatings to act as artificial electrode interfaces in addition to comprising the electrode materials themselves. Newer silicon electrodes enable ultrahigh capacity compared to commercialized graphite anodes[167], but the formation of a solid-electrolyte-interface (SEI) limits the electrochemical performances of silicon anodes. After coating with a lithium-conducting COF thin film as an artificial SEI, the formation of native SEI was suppressed and the rate performance, cycling performance and columbic efficiency of COF coated silicon anode were improved.[168] COF thin films could also be applied to protect lithium anodes. Lithium anodes were considered one of the most promising candidates for high energy density batteries. However, the commercialization of lithium anode has been largely hindered by the undesirable Li dendrite growth, which can cause serious issues including end user safety. Meng et al. and Guo et al. reported that COF thin films could serve as an artificial SEI to protect the lithium anode and suppress dendrite growth.[169,170] The pore structures in COF thin film could serve as channels for conducting lithium ions and redistribute the lithium flux. This led to homogeneous plating/striping process, due to the robust mechanical properties of COF film and therefore could also restrict dendrite growth. As a result, devices were fabricated using COF thin film interfaces with lithium metal-based battery, demonstrating a system with attractive stability.

## 6. Challenges and Opportunities

One of the biggest challenges facing COF thin films is the molecular level control in their synthesis to minimize undesirable defects and contaminants and to achieve precise structural control over a large area. This point also underlies the importance of developing scalable synthesis method for large-scale high-quality COF thin films critically needed in many practical applications. Another challenge is the rational design, fabrication and application of COF thin films based vertical (stacking or coating) or lateral (stitching or joining) heterostructures with other materials since such heterogeneous integrated systems represent the majority of the application scenarios discussed in this review. Equally important is the challenge of assessing and improving the long-term stability and robustness of COF thin films under external mechanical, chemical, optical and magnetic stimuli. With these and potentially more challenges ahead, we would like to point out exciting opportunities by posing the following questions:

1. Can we further develop and expand the linkage chemistry database to improve the control over crystallinity, orientation and thickness at molecular level and over large area?
2. Can we synthetically construct direct COF/2D and 3D materials based vertical and lateral heterostructures with clean interfaces and boundaries towards unique mechanical, electrochemical, electronic, optical, magnetic properties?
3. Can we ubiquitously prove the long-term survivability of different kinds of COF thin films and heterostructures under different service conditions?

It is our strong belief that efforts made to address these questions will have a long-lasting impact on many exciting directions discussed so far.

## 7. Conclusion/Perspectives

The overwhelming majority of existing studies in covalent organic frameworks are focused on the chemistry associated with synthesizing nanostructured powders. Only recently has the materials science community started to utilize the diverse properties accessible through a tailored design of COF materials. The key enabler to many of the future applications discussed is going to be the reliable, reproducible thin films with controllable morphology, orientation, thickness, and crystallinity. Top-down approaches involving attachment of an already synthesized COF powder to a surface are the most achievable currently, but expected strategies for bottom-up growth will allow for more thin film scalability and property control. Moreover, the combination of COFs with other thin film materials such as 2D inorganics, perovskites, electronic semiconductors, magnetic and structural materials will provide more functionality and customization. It is clear that the future of COFs is vast and impactful, but the construction and assembly of thin films to take advantage of the designer properties will be an exciting challenge to overcome.

**Funding Sources**

This research was funded in part by the Air Force Office of Scientific Research under grant number FA9550-20RXCOR057. Authors would also like to acknowledge the Welch Foundation grant C-1716 and the ARL Cooperative Agreement W911NF-18-2-0062

**Author contributions**

All authors contributed to the review, with LB and QF contributing equally to first authorship and JL and NG equally co-corresponding authors.

**Declaration of Competing Interest**

The authors declare that they have no known competing financial interests or personal relationships that could have appeared to influence the work reported in this paper.

**Data availability statement**

Data available by request to the corresponding authors.